# Foreign co-affiliations and performance measurement of universities and the National Academy of Sciences of Ukraine, 2020-2023

**Corresponding author:** Myrolava Hladchenko, Centre for R&D Monitoring (ECOOM), University of Antwerp, Belgium, hladchenkom@gmail.com

## Abstract

This study examines the role of foreign co-affiliations in shaping the research performance of Ukrainian universities and research institutes of the National Academy of Sciences of Ukraine (NASU) before and during Russia's invasion. In 2023, a notable share of high-visibility articles used for institutional assessment was authored by scholars co-affiliated with both Ukrainian and foreign institutions. This share was higher for NASU (17.1%) than for universities (10.6%), reflecting NASU's research-oriented profile and its strong focus on physical sciences & engineering. In 2022–2023, this share increased across all disciplines in both types of institutions. At universities, the strongest growth occurred in social sciences & humanities, while biomedical & health sciences lagged behind. The largest shares of articles with foreign co-affiliations involved institutions in advanced science systems such as Germany and China, as well as neighboring Poland, Czechia, and Slovakia.

Articles with foreign co-affiliations showed citation impact comparable to internationally co-authored articles and substantially outperformed purely domestic publications.

On the one hand, articles with foreign co-affiliations ensure research continuity and enhance the citation visibility of a resource-constrained and war-affected national science system. On the other hand, they distort the measurement of the national and institutional research performance, thereby hindering reform of the national science system. Reliance solely on international links cannot substitute for a strong domestic base. Evaluation frameworks in Ukraine and other countries could be improved through a two-level differentiation, distinguishing first between types of authorship and second among types of foreign co-affiliations, taking into account their duration and the extent of the affiliated scholars' actual contributions.

Although this lies beyond the primary scope of the present study, the findings highlight that the institutional divide between NASU and universities in Ukraine is not justified by research performance.

## 1. Introduction

Russia's full-scale invasion of Ukraine, launched in 2022, has caused extensive human losses and widespread destruction. Between 2021 and 2023, Ukraine's population declined from 41 million to 33.2 million (RBC, 2024). By February 2024, 6.5 million Ukrainians were abroad as refugees, and 1.3 million had been deported to Russia (BBC, 2024). According to the BBC (2024), Ukraine's GDP fell by 29.1% in 2022, accompanied by an inflation rate of 26%.

Under these conditions, policies that support long-term recovery are critical. Creating favourable social and economic conditions is essential to encourage people to remain in or return to Ukraine. In this context, science can play a key role in sustaining a knowledge-based economy (Scientific Committee of the National Board of Ukraine on Science and Technology, 2023). However, the persistence of Soviet- and post-Soviet institutions continues to constrain the development of science and higher education (Krechetova, 2023; Lukash, 2021; Kuzio, 2001; Riabchuk, 2009; Åslund, 2000, 2001; Kudeila, 2012). Although Ukraine dismantled the Soviet planned economy in 1991 and experienced substantial societal and economic transformations, the organisation and funding of science and higher education have remained largely unchanged. The country maintains a division between the National Academy of Sciences of Ukraine (NASU) and higher education institutions, while sectoral academies in medicine, agriculture, and education were also established. NASU research institutes focus on research, whereas universities combine teaching and research functions.

While research institutes outside universities are not unique to Ukraine, for example, Germany maintains similar structures, funding and management practices differ substantially (Clark, 1983). In contrast to many advanced European countries, basic research funding for universities was introduced in Ukraine only in 2026 (Cabinet of Ministers of Ukraine, 2019). In 2022, NASU received 61.2% of total state research funding, whereas only 16.9% was allocated competitively to universities through the Ministry of Education and Science of Ukraine and the National Research Foundation of Ukraine. The remaining funds were distributed among sectoral academies and other research institutions (Ministry of Education and Science of Ukraine, 2024). In contrast, in Germany, universities receive the main share of state research funding (DFG, 2023, 2024; BMFTR, 2024). NASU, as the major player in the Ukrainian science system, emphasises its research excellence, partly measured through research output and citation impact (NASU, 2025; Loktyev, 2019; Popovych, 2018).

Both national and international experts argue that Soviet-style institutions and insufficiently implemented research funding mechanisms hold Ukrainian science back (Bezvershenko & Kolezhuk, 2022). These effects are particularly critical under conditions of Russia's invasion. We should add that in recent times, publications in Scopus- and Web of Science-indexed journals are among the criteria used in state research evaluation for both universities and research institutes (Ministry of Education and Science of Ukraine, 2018; 2024), as well as among publication requirements for doctoral degrees and promotion to professor (Hladchenko, 2023; Hladchenko & Costas, 2025; Nazarovets, 2026).

Previous studies indicate that in contexts of limited domestic research resources, international collaboration and short-term mobility can serve as compensatory mechanisms by enabling access to advanced research infrastructure and facilitating co-authorship with partners in leading scientific countries (Hottenrott & Lawson, 2017; Safaei et al., 2016; Gingras, 2014). Beyond collaboration alone, research organisations may enhance their research performance through articles involving scholars with foreign co-affiliations. Evidence from Saudi Arabia (Bornmann & Bauer, 2015), South Africa (Andrason & van den Brink, 2023), and Chile (Bachelet et al., 2019) shows how institutions that leverage articles authored by scholars with foreign co-affiliations improve their positions in global rankings and increase the eligibility for performance-based funding. In evolving scientific systems, research organisations may also strategically maintain or confer additional affiliations on highly productive scholars, including members of the scientific diaspora based in core scientific systems such as the United States, China, Germany, and the United Kingdom (Hottenrott et al., 2021). While such practices can enhance measured research output, they can also inflate national and institutional performance and hinder reforms aimed at strengthening organisational research capacity. This dynamic may be relevant to Ukrainian science system, where scholars have experienced chronic underfunding alongside inefficient funding allocation (Bezvershenko & Kolezhuk, 2022). Moreover, Russia's invasion has further exacerbated pre-existing structural weaknesses in Ukraine's research system by damaging the research infrastructure, undermining the working and socioeconomic conditions of Ukrainian scholars, and triggering forced academic mobility (Hladchenko, 2025; Nazarovets, 2026a).

Given these systemic and funding constraints, Ukrainian research organisations may rely on foreign co-affiliations to maintain research output and international visibility. Accordingly, this study addresses two research questions:

RQ1. How did the prevalence and characteristics of articles involving foreign co-affiliations change in NASU institutes and universities between 2020–2021 and 2022–2023, encompassing the periods before and during Russia's full-scale invasion of Ukraine?

RQ2. How did articles involving foreign co-affiliations affect the measurement of research performance of NASU institutes and universities across disciplines during these two periods?

To address these questions, the study compares universities and NASU institutes, examines disciplinary variation, analyses the principal countries associated with foreign co-affiliations, and compares patterns before and during the full-scale invasion. This study contributes to the literature in four ways. First, it provides one of the first systematic comparisons of foreign co-affiliations across universities and NASU institutes within the Ukrainian science system during wartime conditions. Second, it demonstrates how the association between foreign co-affiliations and measured research performance varies across scientific disciplines. Third, it extends discussions of core–periphery relations in science by analysing the geographic structure of foreign co-affiliations in relation to citation impact and publication output. Fourth, by comparing the periods before and during Russia's full-scale invasion, it offers evidence on how wartime disruption reshaped internationalised research practices and their bibliometric implications.

By doing so, the study provides new insights into how international linkages can simultaneously both support national science systems and distort research performance in resource-constrained and crisis-affected science systems.

## 2 The system of science and higher education in Ukraine

In 1991, when Ukraine became an independent state, 47,000 researchers were affiliated with the NASU. It inherited a strong tradition in physics established by internationally renowned theoretical physicists such as Nobel laureate L. Landau, A. Akhiezer, and L. Lifshitz. Within NASU, the physical and technical sciences have traditionally accounted for a larger share of researchers and research institutes than other disciplines (Josephon & Egorov, 1997).

After 1991, the preexisting hierarchical governance structure of the academy was preserved, along with Soviet-era institutional leadership that had long-standing ties to the Communist Party leadership. In 1982, 56 % of academicians were party members. They kept their positions mainly as directors of research institutes and department chairs (Josephson & Egorov, 1997). The Board of Academicians (Presidium) runs NASU (National Academy of Sciences of Ukraine, 2017) with funding for the Board determined through the state budget. In 2020, it was 4.4 million euros (Parliament of Ukraine, 2020). The NASU maintained its Soviet-era organisation into disciplinary sections representing all principal domains of fundamental and applied research. NASU consists of 14 research sections organised into the following disciplinary areas: mathematics; information technology; mechanics; physics and astronomy; earth sciences; material sciences; energy; nuclear physics and energy; chemistry; biology, physiology and molecular biology; general biology; economics; history, philosophy and law; and literature, language and art.

In the early 1990s, most higher education institutions were renamed universities and authorised to conduct both fundamental and applied research (Hladchenko et al., 2020). In 2014, the maximum teaching load of university-affiliated academics was reduced from 900 to 600 hours per year to allow more time for research (Hladchenko & Westerheijden, 2021). Universities did not receive basic research funding until 2026, but university-affiliated academics could apply for competitive project-based research funding provided by the Ministry of Education and Science of Ukraine and NRFU.

After 1991, Ukrainian research assessment policies considered publications only in Ukrainian journals. Since 2013, the Ministry of Education and Science of Ukraine has introduced publications in international journals including those indexed in Scopus and Web of Science in different types of publication requirements e.g. doctoral degrees (Ministry of Higher Education and Science, Youth and Sport of Ukraine, 2012), scientific titles of associate professor and professor (Ministry of Education and Science of Ukraine, 2016), research assessment of higher education institutions (Cabinet of Ministers of Ukraine, 2017).

Overall, Ukraine preserved the Soviet model of research funding allocation. As a result, when the Policy Support Unit of Horizon 2020 evaluated the Ukrainian science system in 2017, it stressed that Ukraine needed to end the blind allocation of research funding and recommended distributing 40% of public research funding through a competitive project-based funding system (Horizon 2020 Policy Support Facilities, 2017). Consequently, in 2018, the former State Fund for Fundamental Research (Cabinet of Ministers of Ukraine, 2018) was reorganised into the National Research Foundation of Ukraine, which began issuing competitive research grants in 2020. However, the share of research funding allocated through NRFU remains relatively small, amounting to 3.6% in 2026.

Russia's invasion of Ukraine brought extra challenges to Ukrainian science system, including a deterioration of the economic situation and damage to the research infrastructure, not to mention regular missile attacks and continuous blackouts (UNESCO, 2024; Lutsenko et al., 2025). Russia's full-scale invasion also prompted responses from the international community (Nazarovets &

Teixeira da Silva, 2022), including governments, universities, and funding bodies aimed at providing Ukrainian scholars with fellowships for temporary stays abroad, including initiatives by the US National Academies of Sciences, the European Commission, and the Aleksander von Humboldt Foundation (OECD, 2022).

## 3. Foreign co-affiliations in academia

Foreign co-affiliations offer significant benefits for both individual scholars and research organisations. For individual scholars, they provide access to additional resources, including funding, research infrastructure, and international collaboration networks. These affiliations are particularly valuable for scholars facing limited institutional support or resource constraints, who may seek complementary resources from foreign institutions (Hottenrott & Lawson, 2017; 2022). Lander (2015) argues that co-affiliations may be even more advantageous than co-authorship: while co-authorship allows researchers to overcome infrastructural limitations through collaboration, co-affiliation provides direct and sustained access to otherwise unavailable resources. Empirical studies indicate that co-affiliated scholars tend to achieve higher productivity and impact (Hottenrott et al., 2021; Tong et al., 2020). Co-affiliation with a foreign institution can also enhance a scholar's visibility and prestige, contributing to career advancement (Hottenrott & Lawson, 2022).

Foreign co-affiliations encompass several distinct phenomena. First, they may reflect dual or multiple institutional appointments, where scholars hold positions at two or more institutions simultaneously. Second, they can be mobility-related, indicating visiting or temporary positions abroad (Krieger, 2016; Schiermeier, 2017). Third, foreign co-affiliations may arise during transitional periods, when scholars move from one employer to another and list both institutions in their publications (Yegros et al., 2021). Fourth, they may take the form of diaspora linkages, whereby scholars who migrate to another country retain formal ties with institutions in their country of origin (Meyer, 2001).

Diaspora co-affiliations and related knowledge networks have reshaped perceptions of highly skilled mobility. They have reframed migration from a "brain drain" to a "brain circulation" model, converting the loss of human capital into a remote, yet accessible, networked asset (Meyer et al., 2001; Meyer & Wattiaux, 2006). Scholars who move abroad but maintain active linkages with institutions in their home country, for example, through co-affiliations, joint projects, or co-authorship, can enhance domestic research productivity (Baruffaldi & Landoni, 2012; Langa, 2018). By acting as transnational bridges, they facilitate knowledge transfer, access to international funding and infrastructure, and integration into global collaboration networks. Such diaspora knowledge networks are particularly prominent in STEM disciplines, which are characterised by high levels of international mobility (Miguelez & Temgoua, 2020). However, the extent to which diaspora knowledge networks translate into tangible developmental gains for countries of origin remains debated (Meyer & Wattiaux, 2006; Breschi et al., 2017).

Structural and policy factors also drive the rise of multiple affiliations. Performance-based research funding, research excellence initiatives, and global university rankings incentivise multiple affiliations that enhance institutional research output (Hottenrott et al., 2021). Analyses of publications from 2008–2014 in Germany, Japan, and the United Kingdom across biology, chemistry, and engineering reveal that the proportion of scholars holding multiple affiliations has at least doubled over the period. Foreign co-affiliations constitute roughly a quarter of multiple affiliations, although their prevalence varies considerably across countries (Hottenrott et al., 2021; Tong & Yang, 2023). These co-affiliations most frequently involve institutions in scientifically advanced countries — notably the United States, China, Germany, and the United Kingdom — forming a core-periphery network. Such arrangements enable scientifically peripheral countries to leverage the resources of advanced systems and enhance their research performance. [deleted]

Disciplinary patterns show that foreign co-affiliations are most common in physics and geosciences (Tong et al., 2020). In leading science systems, e.g., high shares are also observed in clinical & life sciences, reflecting the concentration of costly infrastructure and specialised laboratories in these fields (Cronin, 2001). Social sciences exhibit more heterogeneous patterns, with higher prevalence in the United Kingdom, the United States, Germany, and Italy (Tong et al., 2020).

Despite the above-mentioned benefits, foreign co-affiliations raise ethical and policy concerns (Halevi et al., 2023). Strategic co-affiliation practices can distort institutional performance metrics and university rankings, inflating publication counts and citation impact (Gingras, 2014; Hottenrott & Lawson, 2017; Safaei et al., 2016; Huang & Chang, 2018; Sanfilippo et al., 2018). Case studies from Saudi Arabia (Bornmann & Bauer, 2015), South Africa (Andrason & van den Brink, 2023), and Chile (Bachelet et al., 2019) illustrate how institutions have used foreign co-affiliations to improve their positions in global rankings positions or increase eligibility for performance-based funding. While these strategies can enhance short-term performance indicators, they may undermine genuine institutional development and long-term capacity building (Hottenrott & Lawson, 2017**).**

In this work, foreign co-affiliation for Ukrainian scientists and their influence on Ukrainian research is the main topic, referring the reader to the aspect of emigration, temporarily or permanently, e.g., to (Rose et al., 2026).

## 4. Research method

Data for this study were retrieved from the in-house Scopus database at the Centre for Science and Technology Studies (CWTS), Leiden University. First, we extracted all articles published between 2020 and 2023 with at least one author affiliated with a Ukrainian institution. The initial dataset comprised 61,424 articles. After data cleaning, the final dataset included 61,109 articles authored by Ukrainian scholars affiliated with Ukrainian institutions. From this dataset, we retained 57,010 articles authored by scholars affiliated with either universities or institutes of the National Academy of Sciences of Ukraine (NASU), excluding publications associated with other types of Ukrainian institutions (e.g., private companies).

Within the resulting dataset of 57,010 articles, we identified 5,391 articles that included at least one author with simultaneous affiliations to Ukrainian and foreign institutions. These were classified as articles with foreign co-affiliations (FCAF). When an article with foreign co-affiliations also included authors affiliated exclusively with foreign institutions, it was still classified as FCAF, prioritising the presence of dual affiliations. Among the FCAF, 5,096 articles (94.5%) were co-authored solely by scholars affiliated with foreign institutions. The second category, FCAU (internationally co-authored articles without dual affiliations), consists of 12,966 publications. The FCAF and FCAU categories are mutually exclusive. The remaining 38,655 articles were authored exclusively by scholars affiliated with Ukrainian institutions (UKRAF).

Scholars affiliated with both the NASU and at least one university were counted twice: once in the “universities” category and once in the “NASU” category, increasing the total number of FCAF cases to 5,693. In addition, scholars affiliated with more than one foreign country were counted separately for each country of affiliation. In total, 6,271 cases were considered, including 2,536 cases for NASU and 3,735 for universities. Each article was assigned to a disciplinary area based on the main subject field in the CWTS in-house Scopus database, which served as a proxy for discipline. Articles assigned by Scopus to more than one disciplinary area were fractionalized across those areas.

Field-normalised citation impact (FNCI) for each article was calculated using the CWTS in-house Scopus database of March 2025. Citations to each article were divided by the global average number of citations for articles published in the same discipline and year, based on the Scopus subject classification system. This procedure normalises citation counts by both discipline and year of publication. To examine systematic differences in citation impact, two linear regression models were estimated using log-transformed normalised citation impact (log(FNCI + 1)) as the dependent variable. Log transformation was applied to reduce the influence of extreme citation values and account for the highly skewed distribution of citation counts, following established bibliometric practice (Thelwall & Wilson, 2014). Model 1 included institution type, authorship type, and all two- and three-way interaction terms among these variables. Model 2 extended the analysis by incorporating discipline as an additional independent variable. Models were estimated in R. Data visualisation was conducted in R using the ggplot2 package.

## 5. Results

*5.1 Articles authored by scholars with foreign co-affiliations across authorship types and disciplines*

Figure 1 illustrates the distribution of Ukrainian articles (at least one author is affiliated with Ukrainian institute) across three mutually exclusive article types: articles with only Ukrainian affiliations, articles with only authors with Ukrainian affiliation, but at least one of them has a foreign co-affiliation, and articles with at least one author who has only a foreign affiliation (Ukrainian authors may or may not have a foreign co-affiliation). From 2020 to 2023, the share of articles authored by scholars with foreign co-affiliations increased for both types of institutions under study, rising from 12.0% to 17.1% for NASU and from 5.9% to 10.6% for universities, where we note that this period also shows a growth in the absolute number of these articles. In contrast, the share of internationally co-authored articles decreased in 2023 for both types of institutions, with a more pronounced drop for NASU than for universities. Articles authored solely by scholars with Ukrainian affiliations remained the majority for both NASU and universities, with universities had a higher share. However, the gap between universities and NASU decreased over time due to the rising share of articles with foreign co-affiliations among university-based scholars.

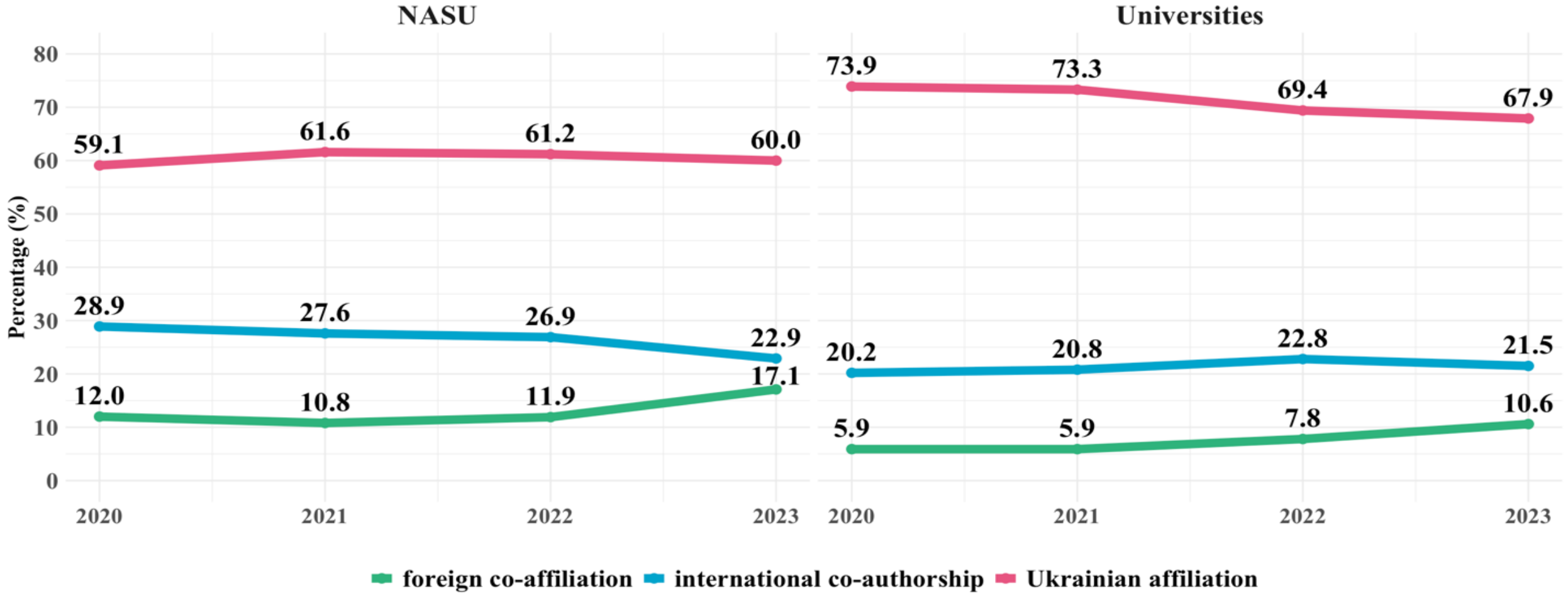

**Fig. 1.** Distribution of articles across authorship types.

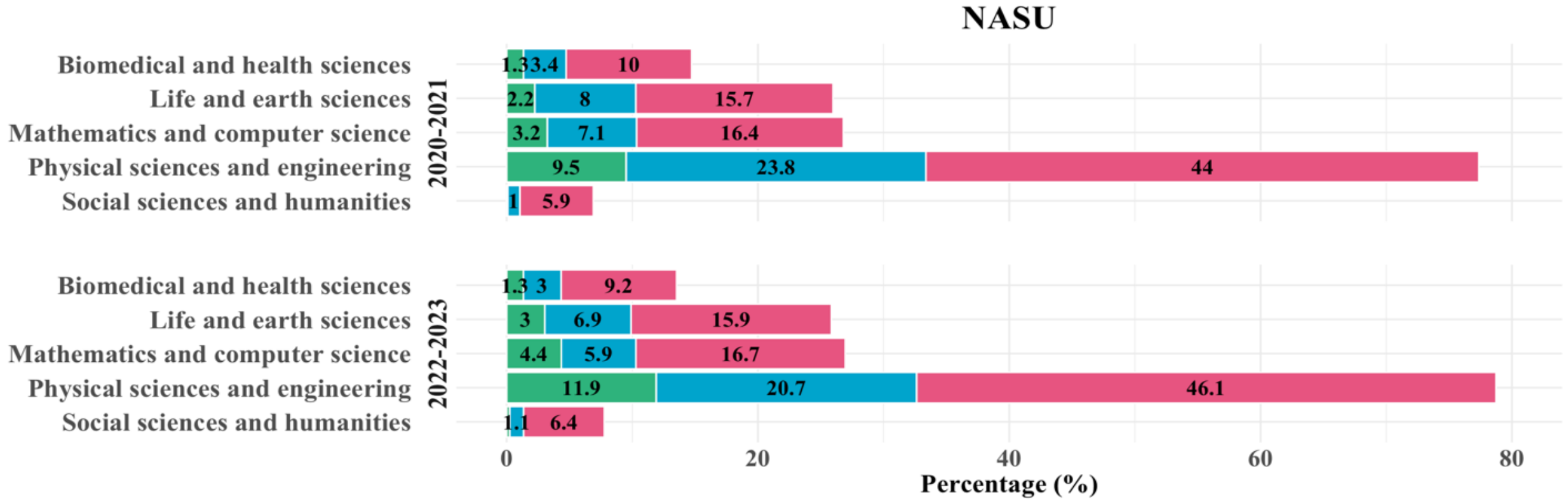

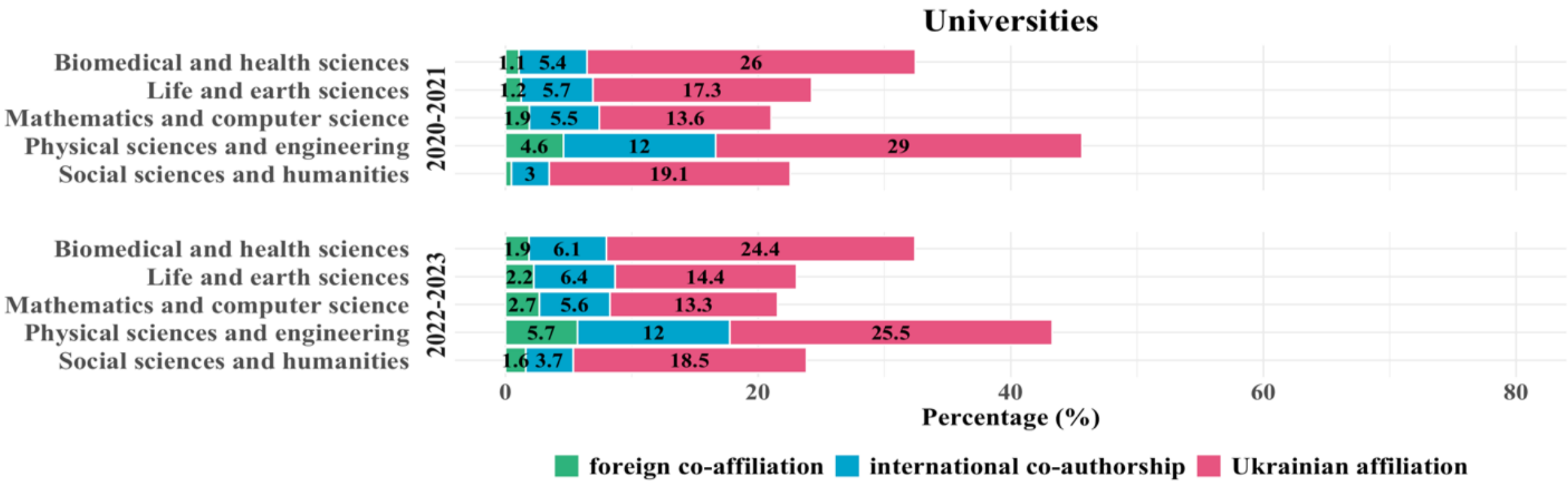

**Fig. 2.** Distribution of articles across authorship types and disciplines (percentage of total per period and type of institution, fractional counting)

Figure 2 shows the distribution of articles across authorship types and disciplines for NASU and universities (percentage of total per period and institution type, fractional counting). For both NASU and universities, the largest share of articles with foreign co-affiliations was in physical sciences & engineering, whereas social sciences & humanities consistently accounted for the smallest share. From 2020–2021 to 2022–2023, the proportion of articles with foreign co-affiliations increased across most disciplines, except for biomedical & health sciences at NASU. In social sciences & humanities, the share remained particularly low in NASU output, at 0.1% and 0.2% for the two periods, respectively. Overall, the contribution of social sciences & humanities to NASU's research output is modest, despite the existence of 31 research institutes in these disciplines. Although scholarly output in the social sciences & humanities is often supplemented by books and local-language journals, Scopus provides substantial coverage of international journals (Sivertsen, 2014).

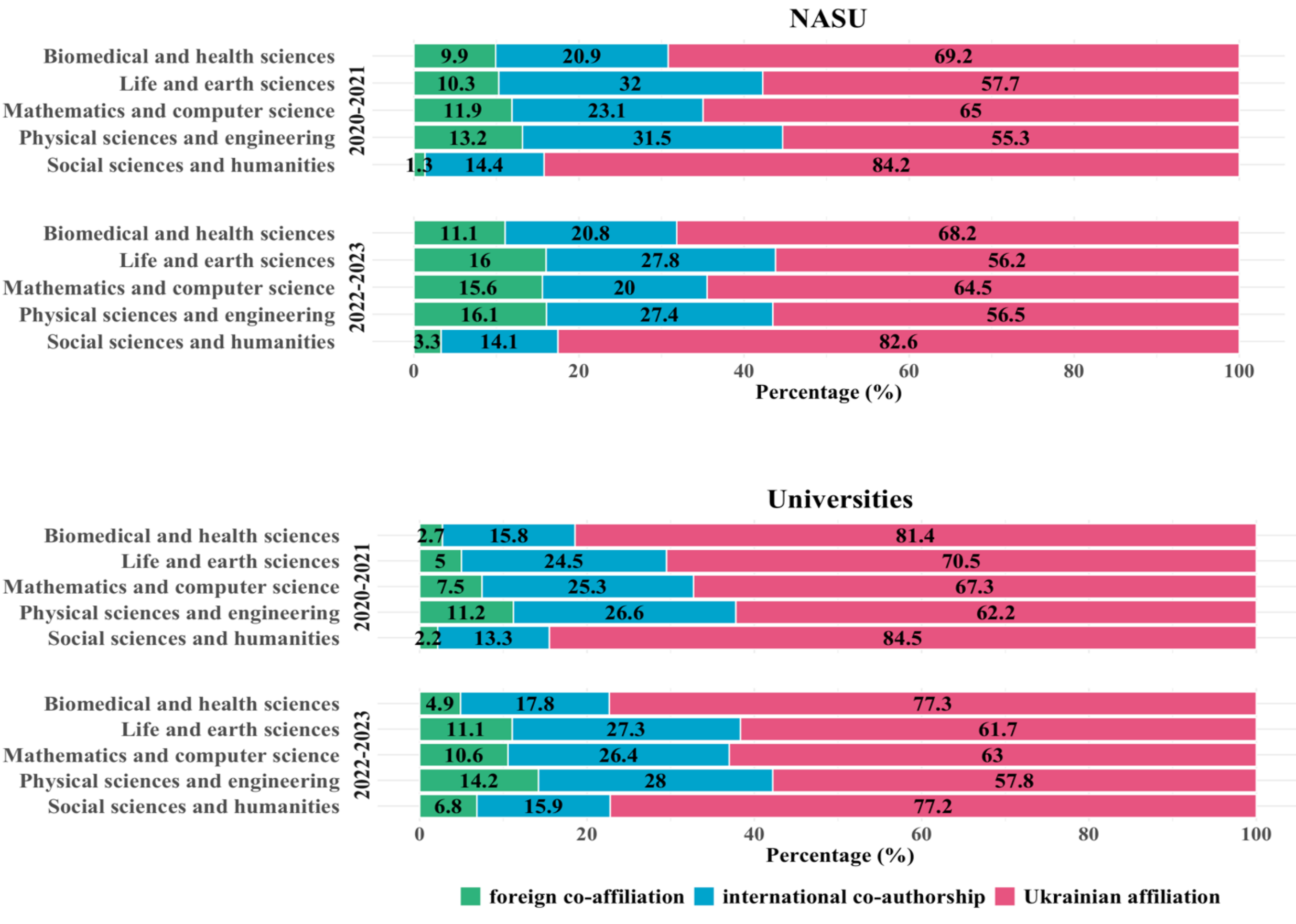

**Fig. 3**. Distribution of articles by authorship type in each discipline (fractional counting).

Figure 3 shows the distribution of articles by authorship type across disciplines. The proportions are calculated so that the total for each discipline equals 100%. For both NASU and universities, the highest share of articles with foreign co-affiliations was observed in physical sciences & engineering, life & earth sciences, and mathematics & computer science in both periods, with NASU consistently exhibiting higher values. The share of articles with foreign co-affiliations increased across all disciplines over time. Notably, in 2022–2023, the share of such articles in the social sciences & humanities (6.8%) at universities exceeded that in the biomedical & health sciences (4.9%).

*5.2 Top ten destinations for foreign co-affiliations*

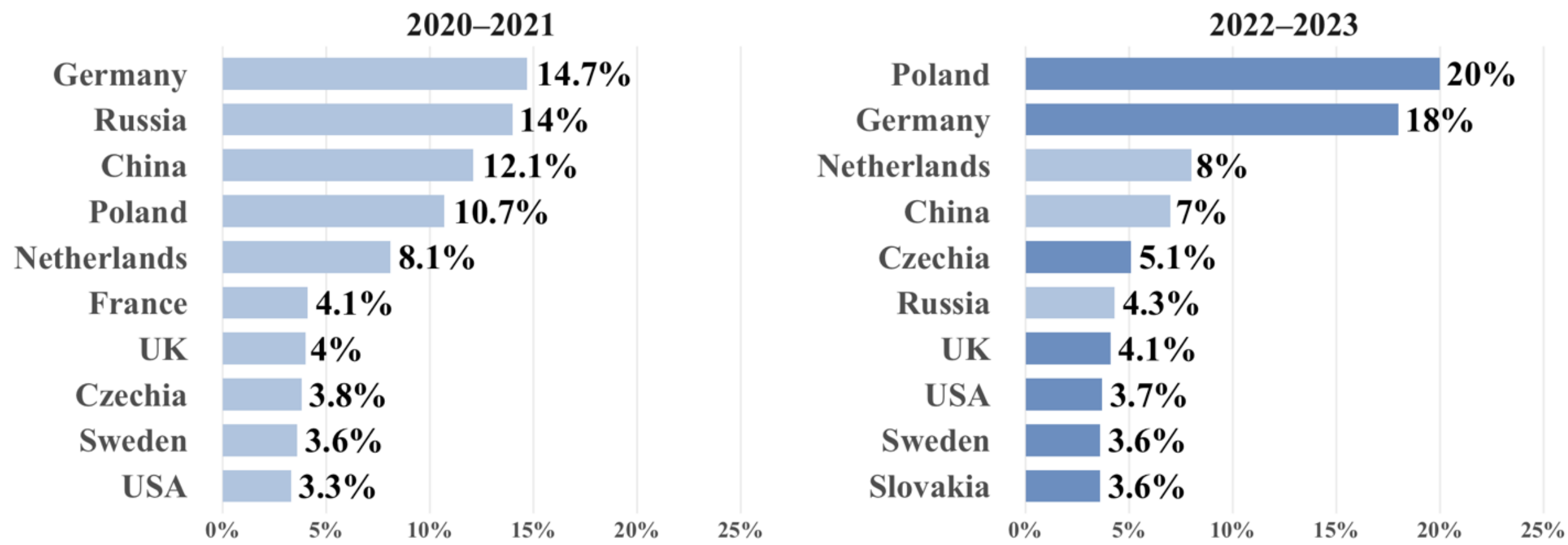

**Fig. 4.** Top ten destinations for foreign co-affiliations (NASU). Increasing shares (right panel) are indicated in a darker blue, while decreasing shares are indicated in a lighter blue.

Table 1. FNCI statistics for articles authored by NASU scholars with foreign co-affiliations across top ten destination countries and periods

| Country | N | Mean FNCI | Median FNCI | SD | SE | 95% CI lower | 95% CI upper | P95 FNCI |
|---|---|---|---|---|---|---|---|---|
| 2020-2021 | | | | | | | | |
| Germany | 162 | 0.84 | 0.51 | 1.09 | 0.09 | 0.68 | 1.01 | 3.1 |
| Russia | 156 | 1.14 | 0.55 | 1.66 | 0.13 | 0.89 | 1.4 | 4.77 |
| China | 133 | 0.7 | 0.43 | 0.77 | 0.07 | 0.57 | 0.83 | 1.97 |
| Poland | 114 | 0.52 | 0.36 | 0.54 | 0.05 | 0.43 | 0.62 | 1.39 |
| Netherlands | 90 | 2.41 | 1.39 | 2.97 | 0.31 | 1.79 | 3.02 | 9.44 |
| UK | 45 | 0.75 | 0.53 | 0.85 | 0.13 | 0.5 | 0.99 | 2.61 |
| France | 44 | 0.9 | 0.51 | 0.98 | 0.15 | 0.62 | 1.19 | 3.06 |
| Czechia | 42 | 0.9 | 0.47 | 2.22 | 0.34 | 0.23 | 1.57 | 2.01 |
| Sweden | 40 | 0.58 | 0.36 | 0.69 | 0.11 | 0.37 | 0.8 | 2 |
| USA | 37 | 0.58 | 0.24 | 0.88 | 0.14 | 0.3 | 0.87 | 2.13 |
| 2022-2023 | | | | | | | | |
| Poland | 279 | 0.84 | 0.34 | 2.98 | 0.15 | 0.55 | 1.13 | 2.71 |
| Germany | 249 | 0.64 | 0.3 | 0.99 | 0.05 | 0.53 | 0.74 | 2.82 |
| Netherlands | 114 | 2.08 | 0.74 | 3.99 | 0.29 | 1.52 | 2.65 | 9.44 |
| China | 99 | 0.68 | 0.45 | 0.81 | 0.06 | 0.57 | 0.79 | 2.58 |
| Czechia | 71 | 1.18 | 0.41 | 2.95 | 0.31 | 0.57 | 1.79 | 4.48 |
| Russia | 59 | 1.33 | 0.59 | 2.42 | 0.28 | 0.79 | 1.87 | 4.78 |
| UK | 57 | 0.49 | 0.17 | 0.74 | 0.08 | 0.33 | 0.64 | 1.72 |
| USA | 51 | 0.66 | 0.44 | 1.01 | 0.12 | 0.43 | 0.9 | 1.62 |
| Slovakia | 51 | 0.58 | 0.15 | 0.88 | 0.11 | 0.36 | 0.8 | 2.06 |
| Sweden | 50 | 0.57 | 0 | 1.28 | 0.16 | 0.27 | 0.87 | 3.23 |

Figure 4 presents the top ten destinations for foreign co-affiliations in NASU publications. Between 2020–2021 and 2022–2023, the composition of leading countries shifted: while Germany, Russia, and China dominated initially, Poland, Germany, and the Netherlands became the most frequent affiliations in the later period. Notably, the share of articles authored by scholars co-affiliated with Poland, Germany, the Czech Republic, and Slovakia increased, with Poland showing the largest growth. For Poland, this increase was accompanied by a rise in mean FNCI from 0.52 to 0.84. The Netherlands maintained the highest mean FNCI in both periods, 2.41 and 2.08, respectively (Table 1).

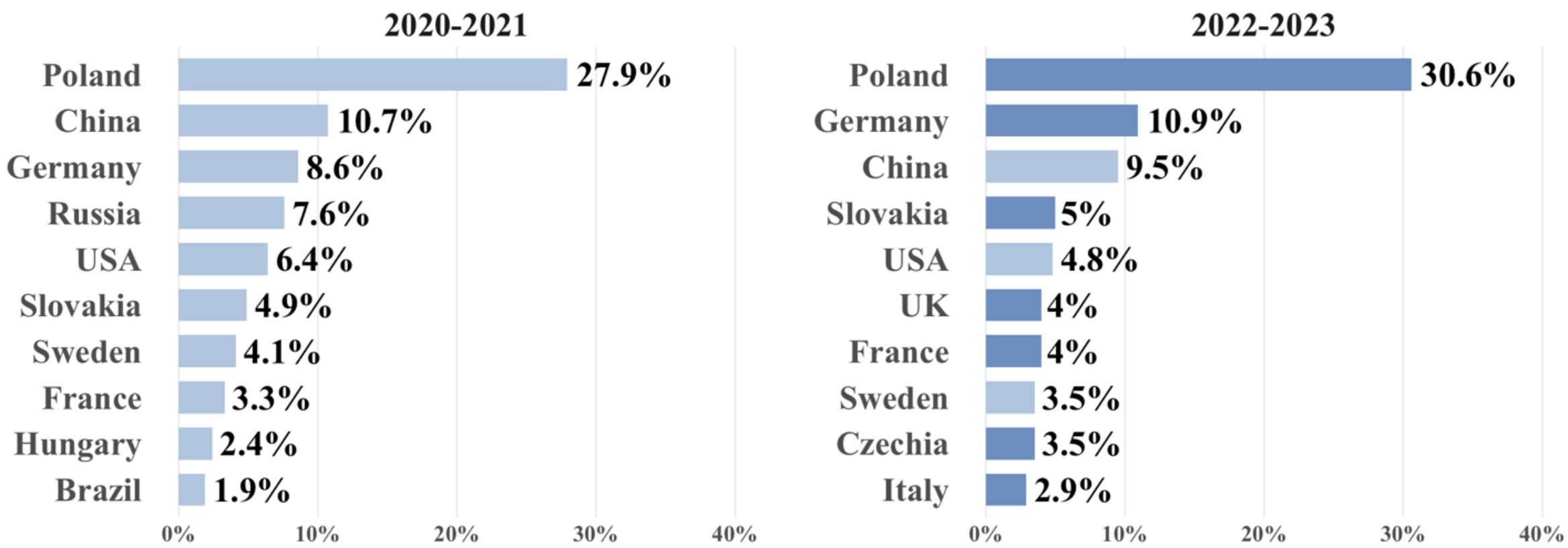

**Fig. 5.** Top ten destinations for foreign co-affiliations (universities).

Table 2. FNCI statistics for articles authored by university scholars with foreign co-affiliations across the top ten destination countries

| Country | N | Mean FNCI | Median FNCI | SD | SE | 95% CI lower | 95% CI upper | P95 FNCI |
|---|---|---|---|---|---|---|---|---|
| 2020-2021 | | | | | | | | |
| Poland | 405 | 0.74 | 0.33 | 1.37 | 0.07 | 0.61 | 0.87 | 2.51 |
| China | 155 | 0.86 | 0.49 | 1.12 | 0.09 | 0.68 | 1.03 | 2.58 |
| Germany | 125 | 0.7 | 0.43 | 0.88 | 0.08 | 0.55 | 0.86 | 2.4 |
| Russia | 110 | 0.42 | 0.19 | 0.6 | 0.06 | 0.31 | 0.53 | 1.49 |
| USA | 93 | 1.01 | 0.37 | 2.79 | 0.29 | 0.44 | 1.57 | 2.55 |
| Slovakia | 71 | 1.3 | 0.43 | 3.36 | 0.40 | 0.52 | 2.08 | 4.49 |
| Sweden | 59 | 0.83 | 0.67 | 0.76 | 0.10 | 0.64 | 1.02 | 2.21 |
| France | 48 | 0.95 | 0.52 | 1.17 | 0.17 | 0.61 | 1.28 | 3.26 |
| Hungary | 35 | 0.47 | 0.43 | 0.38 | 0.06 | 0.35 | 0.6 | 1.1 |
| Brazil | 27 | 0.38 | 0.21 | 0.44 | 0.09 | 0.21 | 0.54 | 1.16 |
| **2022-2023** | | | | | | | | |
| Poland | 697 | 1.39 | 0.42 | 6.54 | 0.25 | 0.9 | 1.88 | 4.44 |
| Germany | 249 | 0.71 | 0.4 | 1.15 | 0.07 | 0.57 | 0.85 | 2.21 |
| China | 216 | 1.08 | 0.49 | 1.81 | 0.12 | 0.84 | 1.32 | 5.14 |
| Slovakia | 114 | 1.18 | 0.5 | 1.85 | 0.17 | 0.84 | 1.52 | 5.03 |
| USA | 109 | 4.25 | 0.43 | 33.3 | 3.19 | -2 | 10.51 | 6 |
| France | 92 | 0.87 | 0.42 | 2.18 | 0.23 | 0.42 | 1.31 | 2.24 |
| UK | 91 | 3.41 | 0.8 | 6.02 | 0.63 | 2.17 | 4.65 | 17.16 |
| Sweden | 81 | 0.69 | 0.4 | 0.87 | 0.10 | 0.5 | 0.88 | 2.24 |
| Czechia | 80 | 0.87 | 0.42 | 1.35 | 0.15 | 0.57 | 1.16 | 3.04 |
| Italy | 67 | 1.51 | 0.74 | 3.53 | 0.43 | 0.66 | 2.35 | 4.1 |

Figure 5 illustrates the top co-affiliation destinations in articles authored by university scholars. Poland consistently ranked first in both periods, with a substantially higher share than NASU (14.7% vs 27.9% in 2020–2021; 20.0% vs 30.6% in 2022–2023). Other countries, including Germany, China, the UK, and Czechia, showed a notable increase in their shares over time. The mean FNCI of articles authored by scholars co-affiliated with Poland rose, from 0.74 to 1.39, exceeding the growth observed for NASU. The mean FNCI also increased for articles authored by scholars co-affiliated with China, the UK and the USA. For Slovakia, the mean FNCI remained above the world average (FNCI = 1) across both periods. (Table 2).

*5.3. Citation impact across authorship types*

Table 3 presents FNCI statistics by institution type, authorship type, and period. In both periods, articles authored solely by scholars with Ukrainian affiliations (UKRAF) exhibit substantially lower mean FNCI compared to internationally co-authored articles and those authored by scholars with foreign co-affiliations.

For both NASU and universities, a pronounced gap between mean and median FNCI is observed for internationally co-authored articles (FCAU) and articles with foreign co-affiliations (FCAF), indicating that mean citation impact is strongly influenced by highly cited outliers. This effect is more pronounced for universities than NASU and contributes to higher mean FNCI, particularly in the second period.

Table 3. FNCI statistics by institution type, authorship type, and period

| Period | Entity | Affiliation | N | Mean FNCI | Median FNCI | SD | SE | 95% CI lower | 95% CI upper | P95 FNCI |
|---|---|---|---|---|---|---|---|---|---|---|
| 2020-2021 | NASU | FCAF | 1033 | 0.91 | 0.48 | 1.45 | 0.05 | 0.82 | 1 | 3.16 |
| | | FCAU | 2550 | 0.79 | 0.38 | 1.31 | 0.03 | 0.74 | 0.84 | 2.82 |
| | | UKRAF | 5453 | 0.2 | 0.1 | 0.39 | 0.01 | 0.19 | 0.22 | 0.79 |
| | | TOTAL | 9036 | 0.45 | 0.17 | 0.95 | 0.01 | 0.43 | 0.47 | 1.71 |
| | Universities | FCAU | 4731 | 1.09 | 0.33 | 10.7 | 0.16 | 0.79 | 1.4 | 2.77 |
| | | FCAF | 1364 | 0.87 | 0.41 | 1.73 | 0.05 | 0.78 | 0.96 | 3.15 |
| | | UKRAF | 16972 | 0.24 | 0.09 | 0.46 | 0.00 | 0.23 | 0.24 | 0.98 |
| | | TOTAL | 23067 | 0.45 | 0.12 | 4.89 | 0.03 | 0.39 | 0.51 | 1.57 |
| 2022-2023 | NASU | FCAF | 1235 | 0.94 | 0.42 | 2.4 | 0.07 | 0.8 | 1.07 | 2.83 |
| | | FCAU | 2118 | 0.83 | 0.42 | 1.36 | 0.03 | 0.77 | 0.89 | 2.91 |
| | | UKRAF | 5159 | 0.25 | 0.11 | 0.5 | 0.01 | 0.23 | 0.26 | 0.99 |
| | | TOTAL | 8512 | 0.49 | 0.17 | 1.24 | 0.01 | 0.47 | 0.52 | 1.86 |
| | Universities | FCAF | 2058 | 1.28 | 0.43 | 8.67 | 0.19 | 0.91 | 1.66 | 3.88 |
| | | FCAU | 4962 | 1.07 | 0.42 | 3.01 | 0.04 | 0.98 | 1.15 | 3.56 |
| | | UKRAF | 15396 | 0.31 | 0.14 | 0.69 | 0.01 | 0.3 | 0.32 | 1.28 |
| | | TOTAL | 22416 | 0.57 | 0.17 | 3.06 | 0.02 | 0.53 | 0.61 | 2.13 |

In 2020–2021, universities slightly exceeded NASU in mean FNCI for internationally co-authored articles (1.09 vs. 0.79), while the mean FNCI for articles with foreign co-affiliations was almost equal (0.87 vs. 0.91). In 2022–2023, universities gained an advantage for articles with foreign co-affiliations as well (1.28 vs. 0.94). Universities also show a higher mean FNCI within each authorship type in 2022-2023 due to more extreme highly cited outliers, indicating more internationally successful research.

Across both periods, the mean FNCI in the TOTAL category increased for both types of institutions, rising from 0.45 to 0.49 for NASU and from 0.45 to 0.57 for universities, indicating a larger increase for universities. However, the median FNCI remained stable at approximately 0.17 for both institution types and both periods, highlighting, that typical articles continued to perform below the world average. The gap between mean and median values, together with higher standard deviations for universities (4.89 and 3.06, compared to 0.95 and 1.24 for NASU), indicates a more unequal distribution of citation impact, driven by a small number of highly cited publications.

Table 4 presents the results of Model 1, a linear regression model predicting log-transformed normalised citation impact (log(FNCI + 1)). The log-transformation was applied to account for the strong right-skewness typical of citation distributions and to reduce the influence of extreme values. The model examines the effects of institution type (university vs NASU), authorship type (FCAF, FCAU, UKRAF), publication period (2020–2021 vs 2022–2023), and their interactions. The model explains approximately 12.7% of the variance (adjusted $R^2 = 0.127$), which is consistent with typical findings in bibliometric analyses, given the high variability in citation counts. Total FNCI values were computed as estimated marginal means from the fitted regression model using the *emmeans* package, averaging over all non-focal covariates included in the model.

To further account for the skewed distribution of citation data, we additionally conducted non-parametric Kruskal–Wallis tests on the original FNCI values. The results indicate statistically significant differences across authorship type ($\chi^2 = 136.57$, $p < 0.001$), publication period ($\chi^2 = 32.22$, $p < 0.001$), and affiliation ($\chi^2 = 8208.6$, $p < 0.001$). These results are broadly consistent with those obtained from the log-transformed regression model, providing additional support for the robustness of the findings.

Relative to the baseline category - university-affiliated UKRAF articles in 2020–2021, articles with foreign co-affiliations (FCAF) exhibited an increase of 0.294 in log(FNCI + 1) ($p < 2e{-}16$), while internationally co-authored articles (FCAU) showed an increase of 0.265 ($p < 2e{-}16$). The period effect was also statistically significant for the baseline category, indicating an increase of 0.035 ($p < 2e{-}16$) in log(FNCI + 1) for UKRAF articles published by universities in 2022–2023.

NASU affiliation was associated with a slight decrease in log(FNCI + 1) (β = −0.014, p = 0.0098). Interactions indicated that NASU FCAF and FCAU articles maintained higher citation gains relative to NASU UKRAF articles. However, the three-way interaction for NASU FCAF in 2022–2023 was negative (−0.081, p = 0.0001), revealing a modest reduction in the log(FNCI + 1) effect relative to the baseline period.

**Table 4** Log-linear model coefficients for the effects of institution type, authorship type, and period on FNCI

| Coefficient | **Estimate** | **Std Error** | **CI lower** | **CI upper** | **p.value** | **Significance** |
|---|---|---|---|---|---|---|
| (Intercept) | 0.173 | 0.003 | 0.17 | 0.18 | < 2e-16 | *** |
| NASU | - 0.014 | 0.006 | -0.03 | -0.003 | 0.0098 | ** |
| 2022-2023 | 0.035 | 0.004 | 0.03 | 0.04 | < 2e-16 | *** |
| FCAF | 0.294 | 0.010 | 0.27 | 0.31 | < 2e-16 | *** |
| FCAU | 0.265 | 0.006 | 0.25 | 0.28 | < 2e-16 | *** |
| NASU:2022-2023 | -0.014 | 0.008 | -0.03 | 0.002 | 0.093 | |
| NASU:FCAF | 0.059 | 0.016 | 0.03 | 0.09 | 0.0002 | *** |
| NASU:FCAU | 0.032 | 0.010 | 0.02 | 0.05 | 0.002 | ** |
| 2022-2023:FCAF | 0.019 | 0.013 | -0.01 | 0.04 | 0.153 | |
| 2022-2023:FCAU | 0.023 | 0.008 | 0.01 | 0.04 | 0.006 | ** |
| NASU:2022-2023:FCAF | -0.081 | 0.021 | -0.12 | -0.04 | 0.0001 | *** |
| NASU:2022-2023:FCAU | -0.027 | 0.015 | -0.06 | 0.003 | 0.080 | |

Significance levels:***p < 0.001,**p < 0.01, *p < 0.05

Table 5 presents predicted FNCI from the log-linear model by institution type, authorship type, and period. It shows consistently higher citation impact for FCAF and FCAU articles across both institution types and periods. UKRAF articles published by universities (0.19 in 2020–2021; 0.23 in 2022–2023) and NASU (0.17 and 0.20) exhibited the lowest and broadly similar predicted FNCI values across both periods.

Internationally co-authored articles (FCAU) showed higher predicted FNCI, increasing from 0.55 to 0.64 for universities and from 0.58 to 0.60 for NASU between the two periods. The predicted FNCI for articles with foreign co-affiliations (FCAF) was comparable to that of internationally co-authored articles, at 0.67 and 0.60 for NASU and 0.59 and 0.68 for universities. Notably, predicted FNCI values for both FCAU and FCAF remained below the world average (1.0). These findings were consistent with Table 3, which showed that the higher mean FNCI in these categories was driven by a small number of highly cited outliers.

Overall, citation impact was broadly comparable between NASU and universities, although universities exhibited moderately higher values in 2022–2023.

**Table 5** Predicted FNCI from log-linear model by institution type, authorship type, and period

| Type of institution | Authorship type | Predicted FNCI | FNCI lower | FNCI upper | Predicted FNCI | FNCI lower | FNCI upper |
|---|---|---|---|---|---|---|---|
| | | 2020-2021 | | | 2022-2023 | | |
| NASU | FCAF | 0.67 | 0.63 | 0.70 | 0.60 | 0.57 | 0.63 |
| | FCAU | 0.58 | 0.55 | 0.60 | 0.60 | 0.58 | 0.63 |
| | UKRAF | 0.17 | 0.16 | 0.18 | 0.20 | 0.19 | 0.21 |
| | TOTAL | 0.45 | 0.44 | 0.48 | 0.45 | 0.44 | 0.48 |
| Universities | FCAF | 0.59 | 0.56 | 0.62 | 0.68 | 0.65 | 0.71 |
| | FCAU | 0.55 | 0.53 | 0.56 | 0.64 | 0.62 | 0.66 |
| | UKRAF | 0.19 | 0.18 | 0.20 | 0.23 | 0.22 | 0.24 |
| | TOTAL | 0.43 | 0.42 | 0.44 | 0.50 | 0.49 | 0.51 |

### *5.4. Citation impact of articles authored by scholars with foreign co-affiliations across disciplines*

Table 6 (Appendix) presents the number of articles, mean, standard deviation, median, interquartile range (IQR), and maximum FNCI values for each combination of institution type, authorship type, discipline, and period. The "TOTAL" category reflects all publications within each group. Articles with foreign co-affiliations (FCAF and FCAU) exhibited higher mean and median FNCI values than domestic articles (UKRAF), indicating greater international citation

impact. NASU and universities showed similar overall patterns, although universities exhibited higher maximum FNCI values, reflecting a small number of highly cited articles. Variability was more pronounced in FCAF articles, as evidenced by larger standard deviation and IQR values.

Regarding articles authored by scholars with foreign co-affiliations, NASU exhibited higher mean FNCI in physical sciences & engineering, reflecting its core disciplinary focus. However, median FNCI values in this field were very similar between institutions (0.48 for NASU and 0.43 for universities in 2020–2021, and 0.42 for both institution types in 2022–2023). This indicates that the higher mean FNCI observed for NASU in this discipline was driven primarily by a concentration of highly cited articles rather than by uniformly higher citation impact across the distribution. Between the two periods, mean FNCI in physical sciences & engineering slightly declined for both NASU (from 0.96 to 0.88) and universities (from 0.76 to 0.71).

Universities exhibited higher mean FNCI in biomedical & health sciences and in social sciences & humanities. In the latter field, median FNCI values were also relatively high, particularly in 2022–2023.

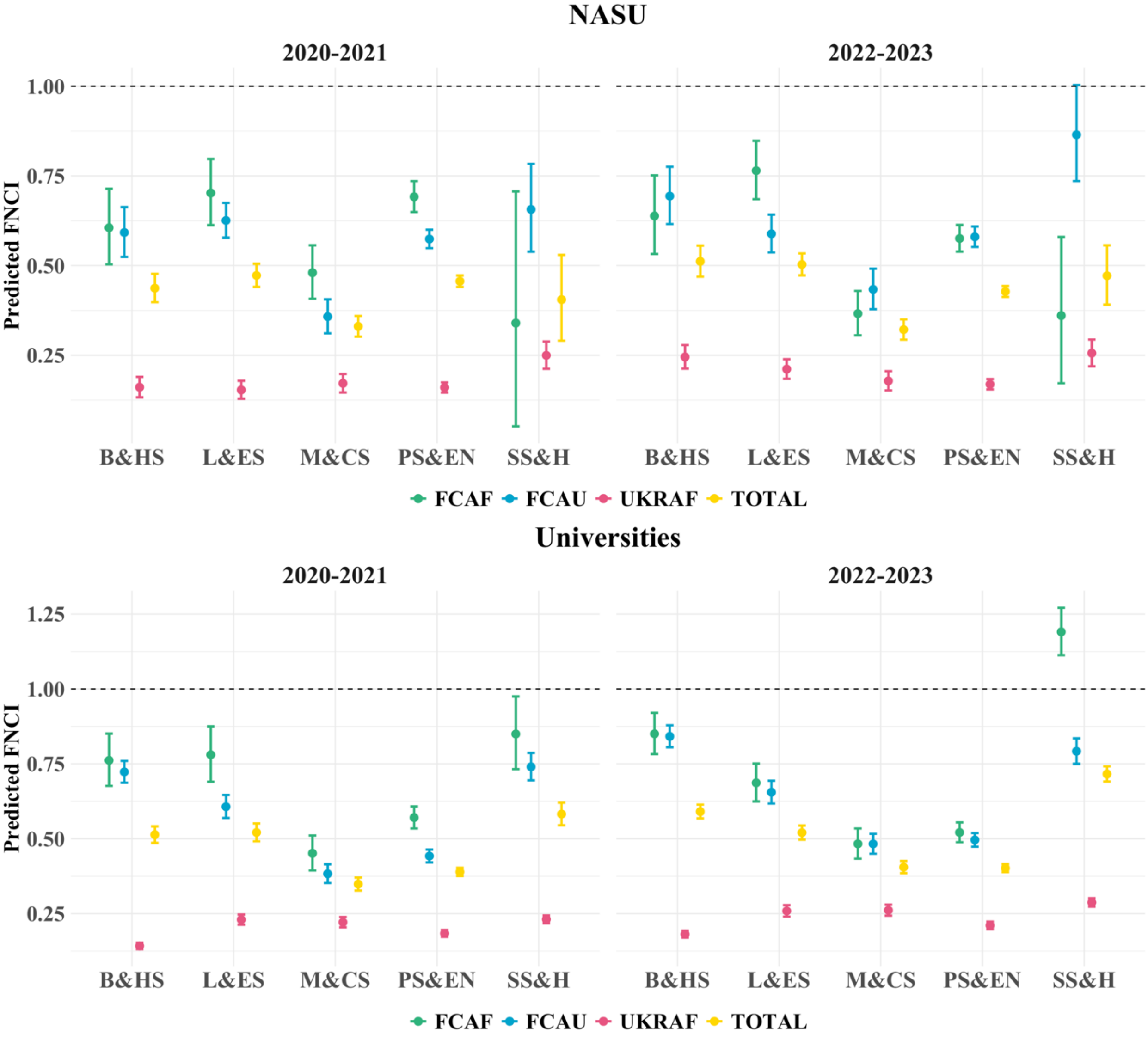

**Fig. 6.** Predicted FNCI from the log-linear model by institution type, authorship type, disciplines, and period.

A linear regression model (Model 2) with interaction terms was fitted to log-transformed citation impact (log(FNCI + 1)) to examine the effects of institutional entity (universities vs. NASU), authorship type, discipline, and all two-, three-, and four-way interactions among these variables. The model explains approximately 14.4% of the variance (adjusted $R^2 = 0.144$). Total FNCI values were computed as estimated marginal means from the fitted regression model using the *emmeans* package, averaging over all non-focal covariates included in the model.

To account for the skewed distribution of citation data, non-parametric Kruskal–Wallis tests were conducted on the original FNCI values. These indicate statistically significant differences across authorship type ($\chi^2$ = 79.38, $p < 0.001$), publication period ($\chi^2$ = 30.03, $p < 0.001$), institutional affiliation ($\chi^2$ = 9231.4, $p < 0.001$), and disciplinary field ($\chi^2$ = 808.3, $p < 0.001$), supporting the robustness of the regression results.

Predicted FNCI values were obtained from the fitted model, back-transformed to the original scale, and are illustrated in Figure 6, while the full results are presented in Table 7 (Appendix). Articles with Ukrainian affiliations (UKRAF) consistently exhibit low relative citation impact across disciplines, with predicted FNCI ranging from 0.14 to 0.29, regardless of period or institution type.

For NASU, FCAF articles showed a predicted citation advantage over FCAU articles in 2020–2021 across all disciplines except social sciences & humanities (SS&H). However, in 2022–2023, this advantage reversed in most disciplines, persisting only in the life & earth sciences. A pronounced decline in predicted FNCI is observed for FCAF articles in mathematics & computer science and in physical sciences & engineering. This shift may reflect changes in the professional profiles of scholars holding foreign co-affiliations in the second period, potentially linked to forced academic mobility.

In 2020–2021, FCAF articles published by universities displayed a modest predicted citation advantage over FCAU articles across all five disciplines. In 2022–2023, this advantage declined slightly in four disciplines but increased markedly in social sciences & humanities (SS&H). Notably, only in social sciences & humanities in 2022–2023 did predicted FNCI exceed 1.0, the world average; in all other categories, it remained below this benchmark. Overall, universities derived the greatest predicted citation advantage from FCAF and FCAU articles in social sciences & humanities, biomedical & health sciences, and life & earth sciences.

For both NASU and universities, predicted FNCI values for FCAF articles had wider confidence intervals than those for FCAU ones, indicating greater variability and uncertainty in citation impact. Taken together, these findings highlight that the citation performance of FCAF articles is more heterogeneous and context-dependent than that of FCAU ones.

Overall, the citation impact of total research output from NASU and universities was broadly comparable across disciplines, with universities generally exhibiting somewhat higher impact, particularly in the social sciences and humanities.

*5.5. Articles in the top 10% most cited globally*

Figure 7 presents the percentage of articles in the top 10% most cited globally, by institution type, authorship type, and period (2020–2023). In 2020, the share of articles in the top 10% most cited globally was higher for NASU than for universities (3.2% vs 2.4%). However, by 2023, the situation had reversed (3.3% vs 4.7%). For universities, growth occurred across all authorship types. Moreover, in 2023, NASU experienced a substantial drop in the share of the top 10% most cited globally in articles with foreign co-affiliations, from 7.8% in 2022 to 5.6% in 2023.

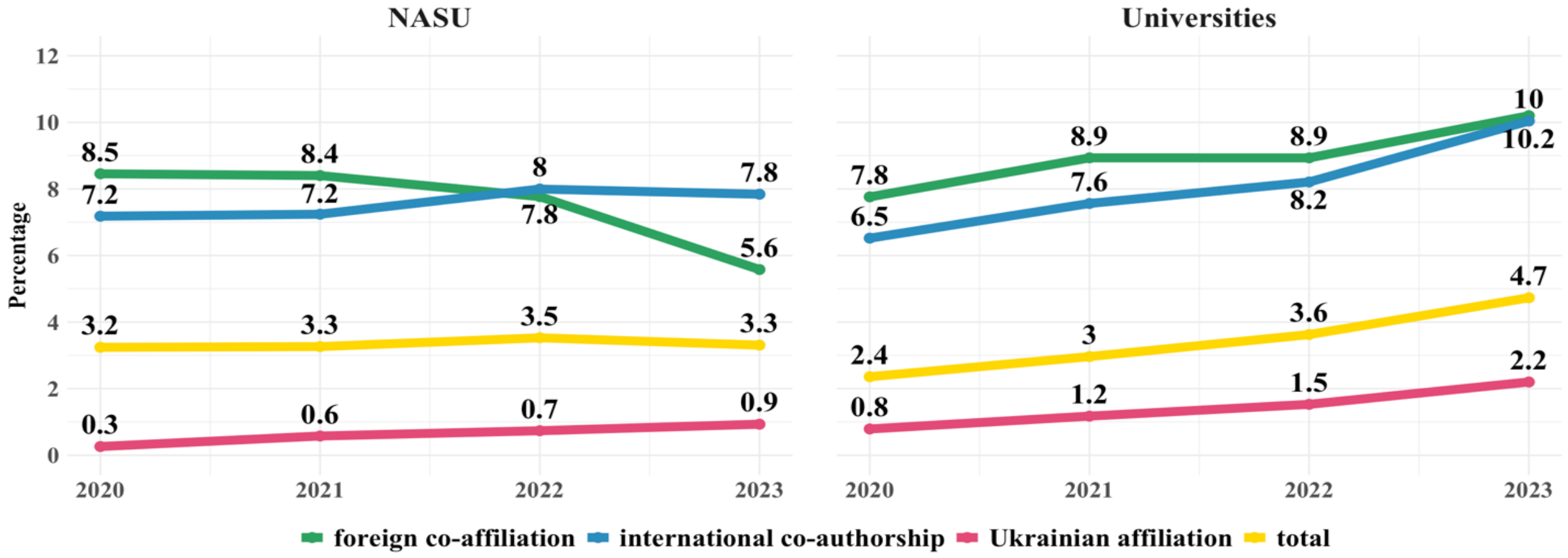

**Fig. 7.** The percentage of articles in the top 10% most cited globally in each category of authorship.

## 6 Discussion & conclusion

### *Foreign co-affiliations and institutional research performance*

The study findings reveal that a substantial share of internationally visible, Scopus-indexed articles, used for institutional research evaluation in Ukraine**,** is authored by scholars with foreign co-affiliations. The proportion of such articles is consistently higher for NASU than for universities, reflecting institutional and disciplinary characteristics. First, the absence of teaching obligations for NASU-affiliated scholars facilitates greater engagement in foreign co-affiliations and research mobility. Second, NASU's profile is dominated by physical sciences & engineering, disciplines that exhibit the highest prevalence of foreign co-affiliations (Tong et al., 2020) and geographical mobility (Hunter et al., 2009; Petersen, 2018; Scellato et al., 2012). Third, these disciplines are highly dependent on advanced research infrastructure, and the limited domestic availability of such infrastructure encourages a compensatory strategy of intensified international engagement, including foreign co-affiliations.

In 2022–2023, the percentage of articles with foreign co-affiliations rose substantially for both NASU and universities. This increase can be associated with the expansion of academic mobility following Russia's full-scale invasion of Ukraine, as documented by Hladchenko (2025). The war simultaneously acted as a push factor through the deterioration of research infrastructure, disrupted working conditions, and the creation of life-threatening circumstances, and as a pull factor by opening new opportunities for international mobility across disciplines, through funding provided by international partners.

### *Disciplinary perspective*

Across both periods, the majority of articles authored by scholars with foreign co-affiliations were concentrated in physical sciences & engineering for both NASU and universities, consistent with Tong et al. (2020). For both institution types, disciplines such as physical sciences & engineering and life & earth sciences consistently had the highest percentage of articles with foreign co-affiliations, reflecting their stronger integration into international scientific networks and higher levels of academic mobility.

In 2022–2023, the proportion of articles with foreign co-affiliations increased across all disciplines for both institution types. For universities, however, the most pronounced rise was in social sciences & humanities, aligning with patterns of international mobility documented by Hladchenko (2025). In contrast, biomedical & health sciences exhibited relatively low representation in both periods, with the lowest share in 2022–2023. This suggests that university-affiliated scholars in this discipline are less integrated into international short- to medium-term mobility networks.

### *Top destinations for foreign co-affiliations*

Analysis of Ukrainian scholars' top destinations for foreign co-affiliations reveals systematic patterns with both advanced and neighbouring science systems, reflecting mobility trends and international research integration. Among advanced systems, Germany and China consistently emerge as leading destinations for foreign co-affiliation for both NASU and universities, confirming prior observations that these countries serve as hubs for international scientific mobility, forming a core-periphery network (Hottenrott et al., 2021; Grogger & Hanson, 2011; Aref et al., 2019). The findings further support Germany's position as the top destination for displaced scholars in Europe, facilitated by extensive third-party funding (Vatansever, 2022). For NASU-affiliated scholars, the Netherlands remained a stable co-affiliation destination across both periods. Although the Netherlands does not provide funding for short-term research visits by Ukrainian scholars, it recruits foreign specialists in the physical sciences due to domestic shortages of physicists (van Hattem, 2011).

Neighbouring countries, primarily Poland, Czechia, and Slovakia, show the influence of geographic and cultural proximity on research collaboration and academic mobility (OECD, 2022; Kiselyova & Ivashchenko, 2024; El-Ouahi et al., 2021; Hladchenko, 2026). Poland's increased prominence corresponds to intensified Ukrainian–Polish collaboration following Russia's 2014 invasion and additional funding to Ukrainian scholars provided by the USA through Poland in 2022 (Research in Poland, 2022).

Taking into account that Russia invaded Ukraine in 2014, it is questionable why there are still articles authored by scholars with co-affiliations to both Ukrainian and Russian institutions, a pattern that remains highly visible in NASU output even in 2022–2023. This issue requires further qualitative investigation.

***Authorship types and citation impact***

The study findings indicate that articles with foreign co-affiliations have citation impact comparable to internationally co-authored articles, whereas both categories substantially outperformed purely domestic publications.

Importantly, although articles with foreign co-affiliations and internationally co-authored articles outperform purely domestic publications, their predicted citation impact generally remains below the world average. This indicates that the observed advantages are relative rather than absolute. The higher mean citation values in these categories are partly driven by a small number of highly cited publications, suggesting that citation gains are unevenly distributed rather than broadly systemic.

Citation benefits from foreign co-affiliations varied across disciplines and institution types. NASU gained a greater citation advantage in biomedical & health sciences, life & earth sciences, and physical sciences & engineering, whereas universities gained a greater citation advantage in biomedical & health sciences, life & earth sciences, and social sciences & humanities.

Following the invasion, the citation gains of articles with foreign co-affiliations decreased for NASU in physical sciences & engineering, likely reflecting the disproportionate impact of the war on disciplines that rely on costly infrastructure. Additionally, highly skilled scholars may have opted for fully foreign affiliations in response to deteriorating local conditions. These findings suggest that maintaining high citation impact through foreign co-affiliations in technologically intensive disciplines requires favourable conditions in the home country; when local conditions deteriorate, highly skilled researchers are more likely to pursue full-time positions abroad or reduce formal ties with institutions in the home country.

In contrast, during the 2022–2023 period, universities saw the emergence of highly cited articles in the social sciences & humanities authored by scholars with foreign co-affiliations, although their overall number remained limited.

***Limitations of the study***

This study has several limitations. First, the analysis focused on articles published in 2022–2023 to capture the effects of Russia’s full-scale invasion of Ukraine. However, many of these articles may have been initiated before the invasion, so that they only partially reflect the impact of the war. Follow-up studies are needed to examine the dynamics of foreign co-affiliations over time and precisely isolate the effects of the invasion.

Second, foreign co-affiliations are operationalised as a single category due to data constraints. In practice, this category encompasses heterogeneous forms of affiliation, including long-term dual appointments, short-term visiting positions, transitional affiliations, and diaspora linkages. These forms likely differ in terms of the intensity of engagement, access to resources, and contribution to research output. Consequently, the results should be interpreted as capturing the aggregate effect of foreign co-affiliations rather than isolating specific underlying mechanisms. This limits the ability to fully assess the institutional implications of foreign co-affiliations. Future research should aim to differentiate between these types to better understand their distinct contributions to research performance.

Third, the operationalisation of articles authored by scholars with foreign co-affiliations may not fully capture their authorship structure, as a substantial share of these articles also has authors affiliated exclusively with foreign institutions. This limits the ability to disentangle the specific role of foreign co-affiliations versus international co-authorship. Future research should refine classification approaches to better distinguish these authorship types.

Fourthly, the analysis did not account for restrictions on male scientific mobility under martial law, adopted after the onset of the invasion. Such constraints may have influenced disciplinary patterns of foreign co-affiliation and should be addressed in future research.

***Reflections and policy directions for Ukraine***

The data reveal structural imbalances in Ukrainian science system, particularly in technologically intensive disciplines such as physical science & engineering and mathematics & computer science, which are critical for Ukraine's resilience, recovery, and reconstruction. Low-cited, domestic articles prevail in these disciplines, highlighting limited local capacity, while articles with foreign co-affiliations and internationally co-authored articles sustain citation visibility. At the same time, the persistently low share of articles with foreign co-affiliations in medical & health sciences signals limited international integration in a societally important domain, where structural constraints, such as limited substitution for teaching duties, reduce opportunities for international mobility for university-affiliated scholars.

On the one hand, articles authored by scholars with foreign co-affiliations ensure research continuity and improve the citation visibility of the national science system, which is constrained by limited resources and affected by war. On the other hand, they distort the measurement of the national and institutional research performance, thereby hindering reform of the national science system. Reliance solely on foreign collaborations cannot substitute for a strong national research base. Building and sustaining local infrastructure, competitive funding mechanisms, and skilled personnel are essential to ensure that international partnerships translate into tangible national benefits. Strengthening domestic capacity in critical disciplines and areas through reform of the science system, including research funding allocation and remuneration policies, will enable Ukraine to meet its resilience, recovery and reconstruction needs and maintain long-term scientific competitiveness.

***Rethinking the institutional divide between NASU and universities***

Although this lies beyond the primary scope of the present study, the findings suggest that the institutional divide between NASU and universities in Ukraine is not justified by research performance. Despite differences in formal missions—where NASU institutes are solely research-focused—the citation impact of their publications is broadly comparable to that of universities across most disciplines, with universities generally exhibiting somewhat higher impact, particularly in the social sciences & humanities. In addition, universities had a higher share of articles in the top 10% most cited globally in 2023.

***Policy implications for research assessment***

First, given that articles with foreign co-affiliations and internationally co-authored articles constitute a substantial share of internationally visible national and institutional research output and demonstrate comparatively high citation impact, they should not be treated as a homogeneous category alongside domestic articles in research evaluation. Accordingly, evaluation frameworks could benefit from differentiating between authorship types in research assessment.

Second, foreign co-affiliations are also not a homogeneous group and include (i) primary employment abroad, (ii) long-term dual appointments, and (iii) diaspora-related affiliations reflecting sustained engagement with domestic institutions. In institutional assessments, outputs might be contextualised according to the type and depth of affiliation, rather than counted uniformly.

The application of such a two-level differentiation—between authorship types and within foreign co-affiliations—would improve the interpretability of performance indicators by distinguishing structurally different forms of authorship and international integration, enabling policymakers to recognise sustained collaboration while limiting its disproportionate influence on research evaluation.

Appendix

**Table 6** Descriptive statistics of FNCI by institution, authorship type, discipline, and period (2020–2021 and 2022–2023)

| Institution | Period | Discipline | Authorship type | N | Mean FNCI | SD FNCI | Median FNCI | IQR FNCI | Max FNCI |
|---|---|---|---|---|---|---|---|---|---|
| NASU | 2020-2021 | Biomedical and health sciences | FCAF | 109 | **0.8** | **1.19** | 0.51 | 0.87 | 8.32 |
| | | | FCAU | 246 | 0.75 | 0.96 | 0.43 | 0.77 | 7.09 |
| | | | UKRAF | 778 | 0.19 | 0.33 | 0.09 | 0.24 | 3.92 |
| | | | TOTAL | 1133 | 0.37 | 0.69 | 0.16 | 0.43 | 8.32 |
| | | Life and earth sciences | FCAF | 160 | **0.89** | **1.32** | **0.6** | 0.93 | 14.58 |
| | | | FCAU | 528 | 0.86 | 1.42 | 0.47 | 0.79 | 14.19 |
| | | | UKRAF | 989 | 0.17 | 0.24 | 0.08 | 0.23 | 2.01 |
| | | | TOTAL | 1677 | 0.46 | 0.98 | 0.18 | 0.42 | 14.58 |
| | | Mathematics and computer science | FCAF | 185 | 0.58 | 0.71 | 0.37 | 0.59 | 4.71 |
| | | | FCAU | 384 | 0.42 | 0.49 | 0.29 | 0.48 | 3.53 |
| | | | UKRAF | 973 | 0.2 | 0.33 | 0.11 | 0.24 | 3.48 |
| | | | TOTAL | 1542 | 0.3 | 0.45 | 0.14 | 0.32 | 4.71 |
| | | Physical sciences and engineering | FCAF | 722 | 0.96 | 1.54 | 0.48 | 0.82 | 13.43 |
| | | | FCAU | 1761 | 0.79 | 1.32 | 0.38 | 0.77 | 21.52 |
| | | | UKRAF | 3157 | 0.19 | 03 | 0.1 | 0.24 | 5.73 |
| | | | TOTAL | 5640 | 0.47 | 1 | 0.18 | 0.43 | 21.52 |
| | | Social sciences and humanities | FCAF | 8 | 0.44 | 0.65 | 0.15 | 0.32 | 1.9 |
| | | | FCAU | 86 | 0.97 | 1.43 | 0.38 | 1.07 | 6.79 |
| | | | UKRAF | 509 | 0.35 | 0.82 | 0.1 | 0.38 | 11.06 |
| | | | TOTAL | 603 | 0.44 | 0.96 | 0.1 | 0.44 | 11.06 |
| | 2022-2023 | Biomedical and health sciences | FCAF | 105 | **0.75** | **0.7** | **0.5** | 0.95 | 3.2 |
| | | | FCAU | 211 | 0.95 | 1.6 | 0.56 | 0.87 | 18.82 |
| | | | UKRAF | 674 | 0.3 | 0.43 | 0.14 | 0.41 | 2.71 |
| | | | TOTAL | 990 | 0.48 | 0.89 | 0.25 | 0.68 | 18.82 |
| | | Life and earth sciences | FCAF | 220 | **1.4** | **4.2** | **0.5** | 0.95 | 50.78 |
| | | | FCAU | 427 | 0.8 | 1.21 | 0.4 | 0.82 | 9.76 |
| | | | UKRAF | 925 | 0.25 | 0.39 | 0.15 | 0.34 | 5.12 |
| | | | TOTAL | 1572 | 0.56 | 1.76 | 0.2 | 0.57 | 50.78 |
| | | Mathematics and computer science | FCAF | 228 | 0.45 | 0.63 | 0.26 | 0.43 | 4.19 |
| | | | FCAU | 302 | 0.57 | 0.91 | 0.32 | 0.61 | 9.09 |
| | | | UKRAF | 905 | 0.23 | 0.51 | 0.1 | 0.28 | 7.12 |
| | | | TOTAL | 1435 | 0.33 | 0.65 | 0.15 | 0.37 | 9.09 |
| | | Physical sciences and engineering | FCAF | 842 | 0.88 | 1.93 | 0.42 | 0.76 | 21.8 |
| | | | FCAU | 1447 | 0.79 | 1.29 | 0.42 | 0.84 | 23.68 |
| | | | UKRAF | 3081 | 0.21 | 0.41 | 0.09 | 0.26 | 9.77 |
| | | | TOTAL | 5370 | 0.47 | 1.1 | 0.17 | 0.51 | 23.68 |
| | | Social sciences and humanities | FCAF | 21 | 0.44 | 0.64 | 0.25 | 0.4 | 2.98 |
| | | | FCAU | 91 | 1.27 | 1.75 | 0.71 | 1.41 | 8.96 |
| | | | UKRAF | 532 | 0.37 | 0.84 | 0.14 | 0.41 | 11.19 |
| | | | TOTAL | 644 | 0.5 | 1.06 | 0.14 | 0.49 | 11.19 |
| Universities | 2020-2021 | Biomedical and health sciences | FCAF | 191 | 1.25 | 2.87 | 0.55 | 0.88 | 26.13 |
| | | | FCAU | 1047 | 2.51 | 22.56 | 0.36 | 0.93 | 673.86 |
| | | | UKRAF | 5255 | 0.18 | 0.41 | 0.06 | 0.18 | 7.9 |
| | | | TOTAL | 6493 | 0.59 | 9.12 | 0.09 | 0.26 | 673.86 |
| | | Life and earth sciences | FCAF | 174 | 1.18 | 2.49 | 0.62 | 0.88 | 26.13 |
| | | | FCAU | 822 | 0.84 | 1.82 | 0.42 | 0.89 | 38.69 |
| | | | UKRAF | 2438 | 0.29 | 0.5 | 0.12 | 0.27 | 7.9 |
| | | | TOTAL | 3434 | 0.46 | 1.17 | 0.18 | 0.43 | 38.69 |
| | | Mathematics and computer science | FCAF | 291 | 0.68 | 1.55 | 0.27 | 0.56 | 19.62 |
| | | | FCAU | 918 | 0.49 | 0.76 | 0.24 | 0.5 | 7.46 |
| | | | UKRAF | 2358 | 0.27 | 0.42 | 0.12 | 0.26 | 4.97 |
| | | | TOTAL | 3567 | 0.36 | 0.69 | 0.14 | 0.34 | 19.62 |
| | | Physical sciences and engineering | FCAF | 857 | 0.76 | 1.25 | 0.43 | 0.72 | 19.62 |
| | | | FCAU | 2120 | 0.58 | 0.96 | 0.29 | 0.58 | 21.52 |
| | | | UKRAF | 5025 | 0.22 | 0.37 | 0.1 | 0.24 | 5.73 |
| | | | TOTAL | 8002 | 0.37 | 0.73 | 0.14 | 0.38 | 21.52 |
| | | Social sciences and humanities | FCAF | 109 | 1.55 | 2.73 | 0.38 | 1.34 | 12.34 |
| | | | FCAU | 679 | 1.06 | 1.54 | 0.48 | 1.37 | 12.15 |
| | | | UKRAF | 4310 | 0.3 | 0.61 | 0.1 | 0.34 | 8.8 |
| | | | TOTAL | 5098 | 0.43 | 0.94 | 0.1 | 0.45 | 12.34 |
| | 2022-2023 | Biomedical and health sciences | FCAF | 338 | 2.66 | 20.8 | 0.57 | 1.12 | 348.03 |
| | | | FCAU | 1178 | 1.7 | 4.64 | 0.49 | 1.29 | 60.39 |
| | | | UKRAF | 4956 | 0.25 | 0.64 | 0 | 0.25 | 11.18 |
| | | | TOTAL | 6472 | 0.64 | 5.23 | 0.14 | 0.45 | 348.03 |
| | | Life and earth sciences | FCAF | 334 | 1.01 | 1.91 | 0.47 | 0.82 | 19.5 |
| | | | FCAU | 888 | 1.09 | 3.74 | 0.45 | 0.83 | 84.03 |
| | | | UKRAF | 1999 | 0.32 | 0.54 | 0.17 | 0.4 | 5.68 |
| | | | TOTAL | 3221 | 0.61 | 2.13 | 0.22 | 0.6 | 84.03 |
| | | Mathematics and computer science | FCAF | 404 | 0.64 | 0.91 | 0.32 | 0.72 | 7.26 |
| | | | FCAU | 927 | 0.64 | 0.95 | 0.32 | 0.7 | 9.73 |
| | | | UKRAF | 2205 | 0.33 | 0.58 | 0.16 | 0.4 | 7.34 |
| | | | TOTAL | 3536 | 0.45 | 0.75 | 0.2 | 0.5 | 9.73 |
| | | | FCAF | 992 | 0.71 | 1.39 | 0.42 | 0.7 | 27.7 |

| | | Physical sciences and engineering | FCAU | 2015 | 0.71 | 2.22 | 0.34 | 0.7 | 84.03 |
|---|---|---|---|---|---|---|---|---|---|
| | | | UKRAF | 4210 | 0.26 | 0.5 | 0.14 | 0.3 | 9.77 |
| | | | TOTAL | 7217 | 0.45 | 1.36 | 0.17 | 0.51 | 84.03 |
| | | Social sciences and humanities | FCAF | 359 | 2.08 | 3.64 | 0.75 | 2.01 | 32.07 |
| | | | FCAU | 835 | 1.22 | 2.12 | 0.49 | 1.28 | 28.31 |
| | | | UKRAF | 4072 | 0.41 | 0.9 | 0.14 | 0.48 | 11.6 |
| | | | TOTAL | 5266 | 0.65 | 1.57 | 0.25 | 0.57 | 32.07 |

Table 7 Predicted FNCI from log-linear model by institution, discipline and period

| **Institution** | **Period** | **Discipline** | **Affiliation type** | **Predicted FNCI** | **FNCI lower** | **FNCI upper** |
|---|---|---|---|---|---|---|
| NASU | 2020-2021 | Biomedical and health sciences | FCAF | 0.61 | 0.50 | 0.71 |
| | | | FCAU | 0.59 | 0.52 | 0.66 |
| | | | UKRAF | 0.16 | 0.13 | 0.19 |
| | | | TOTAL | 0.44 | 0.40 | 0.48 |
| | | Life and earth sciences | FCAF | 0.70 | 0.61 | 0.80 |
| | | | FCAU | 0.63 | 0.58 | 0.67 |
| | | | UKRAF | 0.15 | 0.13 | 0.18 |
| | | | TOTAL | 0.47 | 0.44 | 0.50 |
| | | Mathematics and computer science | FCAF | 0.48 | 0.41 | 0.56 |
| | | | FCAU | 0.36 | 0.31 | 0.41 |
| | | | UKRAF | 0.17 | 0.15 | 0.20 |
| | | | TOTAL | 0.33 | 0.30 | 0.36 |
| | | Physical sciences and engineering | FCAF | 0.69 | 0.65 | 0.74 |
| | | | FCAU | 0.57 | 0.55 | 0.60 |
| | | | UKRAF | 0.16 | 0.15 | 0.17 |
| | | | TOTAL | 0.46 | 0.44 | 0.47 |
| | | Social sciences and humanities | FCAF | 0.34 | 0.05 | 0.71 |
| | | | FCAU | 0.66 | 0.54 | 0.78 |
| | | | UKRAF | 0.25 | 0.21 | 0.29 |
| | | | TOTAL | 0.41 | 0.29 | 0.53 |
| | 2022-2023 | Biomedical and health sciences | FCAF | 0.64 | 0.53 | 0.75 |
| | | | FCAU | 0.69 | 0.62 | 0.78 |
| | | | UKRAF | 0.25 | 0.21 | 0.28 |
| | | | TOTAL | 0.51 | 0.47 | 0.56 |
| | | Life and earth sciences | FCAF | 0.76 | 0.68 | 0.85 |
| | | | FCAU | 0.59 | 0.54 | 0.64 |
| | | | UKRAF | 0.21 | 0.18 | 0.24 |
| | | | TOTAL | 0.50 | 0.47 | 0.53 |
| | | Mathematics and computer science | FCAF | 0.37 | 0.31 | 0.43 |
| | | | FCAU | 0.43 | 0.38 | 0.49 |
| | | | UKRAF | 0.18 | 0.15 | 0.21 |
| | | | TOTAL | 0.32 | 0.29 | 0.35 |
| | | Physical sciences and engineering | FCAF | 0.58 | 0.54 | 0.61 |
| | | | FCAU | 0.58 | 0.55 | 0.61 |
| | | | UKRAF | 0.17 | 0.15 | 0.18 |
| | | | TOTAL | 0.43 | 0.41 | 0.44 |
| | | Social sciences and humanities | FCAF | 0.36 | 0.17 | 0.58 |
| | | | FCAU | 0.86 | 0.74 | 1.00 |
| | | | UKRAF | 0.26 | 0.22 | 0.29 |
| | | | TOTAL | 0.47 | 0.39 | 0.56 |
| Universities | 2020-2021 | Biomedical and health sciences | FCAF | 0.76 | 0.68 | 0.85 |
| | | | FCAU | 0.72 | 0.69 | 0.76 |
| | | | UKRAF | 0.14 | 0.13 | 0.15 |
| | | | TOTAL | 0.51 | 0.49 | 0.54 |
| | | Life and earth sciences | FCAF | 0.78 | 0.69 | 0.88 |
| | | | FCAU | 0.61 | 0.57 | 0.65 |
| | | | UKRAF | 0.23 | 0.21 | 0.25 |
| | | | TOTAL | 0.52 | 0.49 | 0.55 |
| | | | FCAF | 0.45 | 0.39 | 0.51 |
| | | Mathematics and computer science | FCAU | 0.38 | 0.35 | 0.41 |
| | | | UKRAF | 0.22 | 0.20 | 0.24 |
| | | | TOTAL | 0.35 | 0.33 | 0.37 |
| | | Physical sciences and engineering | FCAF | 0.57 | 0.53 | 0.61 |
| | | | FCAU | 0.44 | 0.42 | 0.46 |
| | | | UKRAF | 0.18 | 0.17 | 0.20 |
| | | | TOTAL | 0.39 | 0.38 | 0.40 |
| | | Social sciences and humanities | FCAF | 0.85 | 0.73 | 0.97 |
| | | | FCAU | 0.74 | 0.70 | 0.79 |
| | | | UKRAF | 0.23 | 0.22 | 0.24 |
| | | | TOTAL | 0.58 | 0.55 | 0.62 |
| | 2022-2023 | Biomedical and health sciences | FCAF | 0.85 | 0.78 | 0.92 |
| | | | FCAU | 0.84 | 0.81 | 0.88 |
| | | | UKRAF | 0.18 | 0.17 | 0.19 |
| | | | TOTAL | 0.59 | 0.57 | 0.61 |
| | | Life and earth sciences | FCAF | 0.69 | 0.62 | 0.75 |
| | | | FCAU | 0.66 | 0.62 | 0.69 |
| | | | UKRAF | 0.26 | 0.24 | 0.28 |
| | | | TOTAL | 0.52 | 0.50 | 0.54 |
| | | Mathematics and computer science | FCAF | 0.48 | 0.43 | 0.53 |
| | | | FCAU | 0.48 | 0.45 | 0.52 |
| | | | UKRAF | 0.26 | 0.24 | 0.28 |
| | | | TOTAL | 0.41 | 0.39 | 0.43 |
| | | Physical sciences and engineering | FCAF | 0.52 | 0.49 | 0.55 |
| | | | FCAU | 0.50 | 0.47 | 0.52 |
| | | | UKRAF | 0.21 | 0.20 | 0.22 |
| | | | TOTAL | 0.40 | 0.39 | 0.42 |
| | | Social sciences and humanities | FCAF | 1.19 | 1.11 | 1.27 |
| | | | FCAU | 0.79 | 0.75 | 0.84 |
| | | | UKRAF | 0.29 | 0.27 | 0.30 |
| | | | TOTAL | 0.72 | 0.69 | 0.74 |